\documentclass[runningheads,a4paper]{llncs}

\usepackage{amsmath}
\usepackage{amssymb}
\usepackage{graphicx}
\usepackage{booktabs}
\usepackage[flushleft]{threeparttable}
\usepackage{array}
\usepackage[font=small, labelfont=bf]{caption}
\usepackage{subcaption}
\usepackage[caption=false]{subfig}

\usepackage{url}
\urldef{\mailsa}\path|paul.verschure@upf.edu|    
\newcommand{\keywords}[1]{\par\addvspace\baselineskip
\noindent\keywordname\enspace\ignorespaces#1}

\graphicspath{ {Images/} }

\begin{document}

\mainmatter  

\title{Scaling Properties of Human Brain \\Functional Networks}
\titlerunning{Scaling of Brain Functional Networks}

%
%
\author{Riccardo Zucca\inst{1} \and Xerxes D.\,Arsiwalla\inst{1} \and Hoang Le\inst{2}  \and \\Mikail Rubinov\inst{3,4} \and Paul F.\,M.\,J.\,Verschure\inst{1,5}}
\authorrunning{R. Zucca, X.D. Arsiwalla et al.}

\institute{Laboratory of Synthetic Perceptive, Emotive and Cognitive Systems (SPECS), N-RAS, DTIC, Universitat Pompeu Fabra (UPF), Barcelona, Spain\\ \and California Institute of Technology, Pasadena, USA \and Department of Psychiatry, Behavioural and Clinical Neuroscience Institute, University of Cambridge, Cambridge, UK \and Janelia Research Campus, Howard Hughes Medical Institute, Ashburn, VA, USA \and  Catalan Institute of Advanced Studies (ICREA), Barcelona, Spain
\mailsa\\
\url{http://specs.upf.edu}}

%
%

\toctitle{Scaling Properties of Human Brain Functional Networks}
\tocauthor{Riccardo Zucca, Xerxes D. Arsiwalla, Hoang Le, Mikail Rubinov, Paul F. M. J. Verschure}
\maketitle

\begin{abstract}
We investigate scaling properties of human brain functional networks in the  resting-state. Analyzing network degree distributions, we statistically test  whether their tails scale as power-law or not. Initial studies, based on least-squares fitting, were shown to be inadequate for precise estimation of power-law distributions. Subsequently, methods based on maximum-likelihood estimators have been proposed and applied to address this question. Nevertheless, no clear consensus has emerged, mainly because results have shown substantial variability depending on the data-set used or its resolution. In this study, we work with high-resolution data (10K nodes) from the Human Connectome Project and take into account network weights. We test for the power-law, exponential, log-normal and generalized Pareto distributions. Our results show that the statistics generally do not  support a power-law, but instead these degree distributions tend towards the thin-tail limit of the generalized Pareto model. This may have implications for the number of hubs in human brain functional networks. 
\end{abstract}

\section{Introduction}

Much interest in theoretical neuroscience has revolved around graph-theoretic scaling properties of the network of structural and functional correlations in the human brain. Some authors have described the degree distribution of nodes in brain functional networks as scale-free; that is, these networks follow a power-law degree distribution $P(k) \sim k^{-\alpha}$  with an exponent close to~2  \cite{Eguiluz2005,VandenHeuvel2008}, indicating the presence of a small  number of hub-nodes that connect widely across the network. Other studies have suggested that functional brain networks are not scale-free, but instead are characterized by an exponentially truncated distribution \cite{Achard2006,Fornito2010,Hayasaka2010}. The scaling characteristics of these networks are associated with the number and organization of network hubs and consequently may have implications with our understanding of how the brain responds to disease or damage \cite{Albert2000,Achard2006}. This calls for a rigorous statistical methodology to infer underlying models that best describe the degree distribution of brain functional networks. 

Initial studies were based on least-square fitting of log-log plots of frequency distributions to answer this question. This approach, although seemingly straightforward, is inadequate from a statistical point of view as elaborated in \cite{Clauset2009}. Least-square fitting may give systematically biased estimates of the scaling parameters and most of the inferential assumptions for regression are violated. Moreover, in all these studies no statistical testing was mentioned to measure the \emph{goodness-of-fit} of each fitted degree distribution. As an alternative, Maximum Likelihood Estimation (MLE) of the scaling parameters should be used and alternative distributions should be also tested. In \cite{Clauset2009} an analytical framework for performing such tests for power-law models is provided, which  has been subsequently extended for testing other distributions as well. 
However, it has been noted that results are still very much dependent on the way the data is preprocessed, how the network is extracted, its dimensions and whether one uses region or voxel-based networks. For instance, Hayasaka et al. \cite{Hayasaka2010} found that, although degree distributions of all analysed functional networks followed an exponentially truncated model, the higher the resolution, the closer the distribution was to a power-law.  

In this work, as a first step to address this issue we analyzed the resting-state fMRI (rs-fMRI) data of 10 subjects obtained from the Human Connectome Project database. Using the MLE method, advocated in \cite{Clauset2009}, we estimate the scaling parameters for the best possible fit for a model distribution and then check the \emph{goodness-of-fit} for this distribution by comparing it to synthetic generated data. We do this for four model distributions: power-law, exponential, log-normal and generalized Pareto. The reason for choosing the generalized Pareto model is due to the fact that it  interpolates between fat-tail and thin-tail distributions, including the power-law and exponential as special cases. In what follows, we find that at a resolution of 10K nodes, the statistics favor the generalized Pareto thin-tail distributions.

\section{Materials and Methods}
\subsection{Subjects, imaging data and network extraction}
\begin{figure}[t!]
\centering
\includegraphics[scale=1]{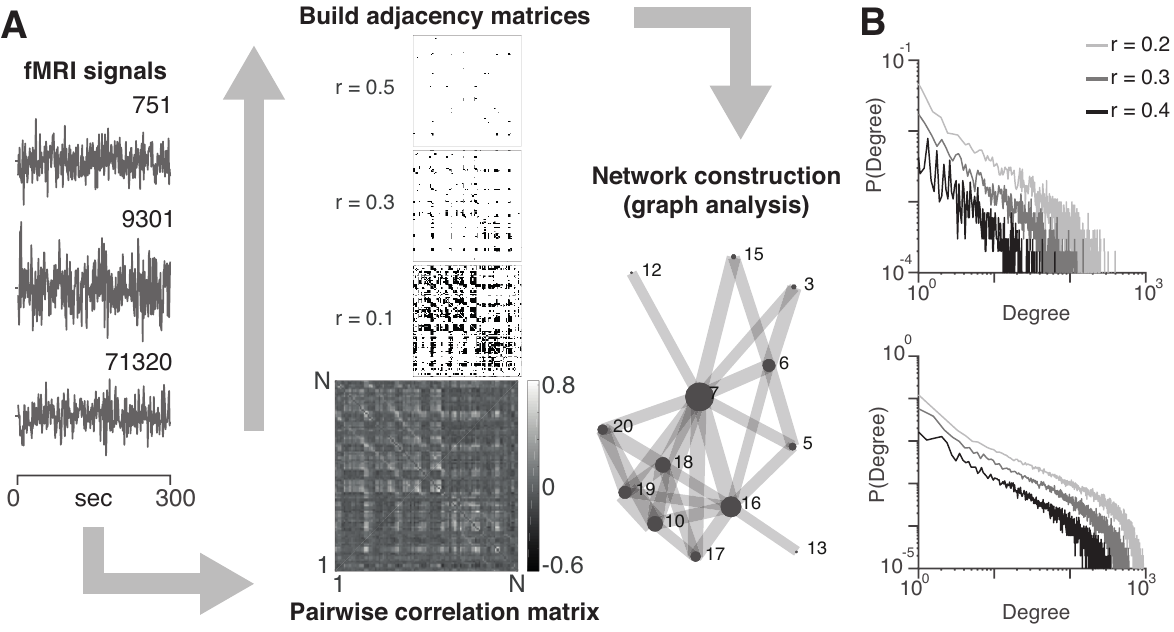}
\caption{\textbf{A.} Overview of the processing steps used to generate graph-based brain connectivity functional networks (see the main text for further details about the overall procedures). \textbf{B.} Degree distributions for three different values of the functional correlation threshold for a representative subject (top) and averaged over the 10 data-sets included in the study (bottom). \label{fig:Protocol}}
\end{figure}

High-quality, high-resolution resting state fMRI scans of 10 subjects from the Human Connectome Project (Q1 data released by the WU-Minn HCP consortium in March 2013 \cite{vanEssen2013}) were analysed in this study (age range: 26--35, 16.7\% male). Individual rs-fMRI data were acquired for  $\sim$15 minutes providing a total of  $\sim$98,300 grayordinates time--series of 1,200 time points each. A schematic illustration of the process used to build the networks is provided in figure~\ref{fig:Protocol}A. Building and visualizing functional networks was done using the BrainX$^3$ platform~\cite{ECMS,2015network,2015connectomics}. For all the subjects, the original data-set was downsampled to $\sim$10,000 nodes by averaging the time-series of neighbouring grayordinates within a 5 mm$^3$ cube. 

Pearson's correlation coefficients were calculated between each possible pair of nodes to build a $N\times N$ functional connectivity matrix, which is symmetric by construction and with self-connections set to zero. The matrix was then thresholded to derive weighted undirected adjacency matrices. 
We examined a range of 18 different thresholds (R) between $-0.7$ and $0.8$, at $0.1$ steps. Outside this range, the functional matrices become too sparse for meaningful analysis. For a positive threshold, each entry in the correlation matrix is set to 0 if its value is less than the threshold value and maintains its value otherwise. For a negative threshold, absolute values of the entries less than the threshold are maintained, while others are set to 0.
In a weighted network, the weighted degree of a node is defined as the sum of all weighted edges connected to that node.  
Figure \ref{fig:Protocol}B illustrates the degree distributions of  extracted networks across three different thresholds for a representative subject and averaged over all 10 data-sets.

\subsection{Fitting parametric models to weighted degree networks\label{methods}}
For every network generated from subject data, the vector of degrees $\textbf{x} = [x_{1}, x_{2}, ...,x_{n}]$ is sorted in ascending order for each  threshold.  For every $x_{i}$, following \cite{Clauset2009}, we use the method of maximum likelihood to estimate the scaling parameter $\alpha$ providing the best possible fit for the hypothetical power-law distribution $P(x)\sim Cx^{-\alpha}$ for the tail of the observed data in the range  $x_{i}$ to $x_{n}$. Next, we calculate the Kolmogorov-Smirnov (KS) statistic for this power-law distribution with respect to $x_{i}$. Out of all possible $x_{i}$ from the data, the one with the smallest KS statistic corresponds to the lower bound $x_{min}$ for power-law behavior in the data. The next step is to verify whether this is indeed a good fit for the data. For that, a large number of synthetic data-sets are sampled from a true power-law distribution with the same scaling parameter $\alpha$ and bound $x_{min}$ as the ones estimated for the best fit of the empirical data.  We fit each synthetic data-set to its own power-law model and calculate the KS statistic for each one relative to its own model. An empirical \emph{p}-value is then calculated as the fraction of the time the empirical distribution outperforms the synthetically generated ones (by having a smaller KS statistic value). If \emph{p}-value $\leqslant 0.1$, the power-law hypothesis can be ruled out as a non plausible explanation of the data. Nevertheless, large \emph{p}-values do not guarantee that the power-law is the best model and the power-law fit has to be compared to a class of competing distributions. 

\begin{table}[t!]   
\centering
\caption{Fit results of the exponential ($p(x)=Ce^{-\lambda x}$) and the power law distributions ($p(x)=Cax^{-\alpha}$)}
\begin{minipage}[t]{0.45\textwidth}
\centering
\tiny
\caption*{Exponential}
\begin{tabular}{lrrrrcc}
\toprule
\multicolumn{1}{c}{Thr} & \multicolumn{1}{c}{$\lambda$} & \multicolumn{1}{c}{$x_{min}$}  & \multicolumn{1}{c}{TL} & \multicolumn{1}{c}{KS} & \multicolumn{1}{c}{p}-value & \multicolumn{1}{c}{TL$_{r}$}\\
\midrule
$+0.8$ & 0.000 & 0.000 & 0.0 & 0.000 & 0.0000 & 0.714 \\   
$+0.7$ & 0.162 & 1.186 & 83.5 & 0.056 & 0.3125 & 0.500 \\  
$+0.6$ & 0.089 & 3.887 & 130.5 & 0.058 & 0.4840 & 0.306 \\ 
$+0.5$ & 0.038 & 15.544 & 175.5 & 0.050 & 0.6170 & 0.187 \\
$+0.4$ & 0.020 & 23.051 & 440.5 & 0.037 & 0.7305 & 0.160 \\ 
$+0.3$ & 0.012 & 44.067 & 805.0 & 0.030 & 0.3305 & 0.167 \\
$+0.2$ & 0.008 & 44.049 & 1552.5 & 0.024 & 0.0895 & 0.205\\
$+0.1$ & 0.006 & 25.520 & 4350.5 & 0.031 & 0.0015 & 0.433\\ 
$+0.0$ & 0.007 & 545.110 & 1597.0 & 0.037 & 0.0290 & 0.159 \\ 
$-0.0$ & 0.021 & 123.255 & 1678.0 & 0.021 & 0.0485 & 0.168 \\ 
$-0.1$ & 0.023 & 53.151 & 141.0 & 0.072 & 0.1835 & 0.014\\   
$-0.2$ & 0.095 & 7.104 & 123.0 & 0.170 & 0.0170 & 0.105\\ 
$-0.3$ & 0.348 & 0.330 & 95.5 & 0.224 & 0.0000 & 0.534 \\  
$-0.4$ & 0.000 & 0.000 & 0.0 & 0.000 & 0.0000 & 0.876\\
$-0.5$ & 0.000 & 0.000 & 0.0 & 0.000 & 0.0000 & 0.881\\
$-0.6$ & 0.000 & 0.000 & 0.0 & 0.000 & 0.0000 & 0.510\\
$-0.7$ & 0.000 & 0.000 & 0.0 & 0.000 & 0.0000 & - \\  
$-0.8$ & 0.000 & 0.000 & 0.0 & 0.000 & 0.0000 & - \\  
\bottomrule
\end{tabular} 
\label{table:EXP}
\end{minipage}
\begin{minipage}[t]{0.45\textwidth}
\centering
\tiny
\caption*{Power law}
\begin{tabular}{lrrrrcc}
\toprule
\multicolumn{1}{c}{Thr} & \multicolumn{1}{c}{$\alpha$} & \multicolumn{1}{c}{$x_{min}$} & \multicolumn{1}{c}{TL} & \multicolumn{1}{c}{KS} &  \multicolumn{1}{c}{p}-value & \multicolumn{1}{c}{TL$_r$}\\
\midrule
$+0.8$ & 0.000 & 0.000 & 0.0 & 0.000 & 0.0000 & 0.655 \\
$+0.7$ & 2.257 & 1.427 & 83.0 & 0.120 & 0.0005 & 0.437 \\
$+0.6$ & 2.324 & 4.387 & 94.5 & 0.099 & 0.0070 & 0.336 \\       
$+0.5$ & 2.765 & 21.077 & 201.0 & 0.085 & 0.0275 & 0.141 \\                   
$+0.4$ & 2.946 & 48.800 & 300.0 & 0.079 & 0.0020 & 0.099 \\                         
$+0.3$ & 3.711 & 180.624 & 224.0 & 0.079 & 0.0650 & 0.045 \\                     
$+0.2$ & 5.541 & 360.395 & 281.5 & 0.079 & 0.0060 & 0.036 \\          
$+0.1$ & 8.615 & 745.120 & 208.0 & 0.0666 & 0.1490 & 0.021 \\                 
$+0.0$ & 12.594 & 931.660 & 246.5 & 0.059 & 0.1810 & 0.025 \\                         
$-0.0$ & 5.303 & 142.645 & 1249.0 & 0.021 & 0.5085 & 0.124 \\      
$-0.1$ & 2.402 & 11.896 & 2352.5 & 0.044 & 0.0000 & 0.236 \\    
$-0.2$ & 2.417 & 3.349 & 521.5 & 0.052 & 0.0030 & 0.277\\         
$-0.3$ & 2.367 & 1.089 & 118.5 & 0.088 & 0.0105 & 0.319\\       
$-0.4$ & 0.000 & 0.000 & 0.0 & 0.000 & 0.0000 & 0.332 \\           
$-0.5$ & 0.000 & 0.000 & 0.0 & 0.000 & 0.0000 & 0.399 \\ 
$-0.6$ & 0.000 & 0.000 & 0.0 & 0.000 & 0.0000 & 0.510 \\    
$-0.7$ & 0.000 & 0.000 & 0.0 & 0.000 & 0.0000 & -\\  
$-0.8$ & 0.000 & 0.000 & 0.0 & 0.000 & 0.0000 & - \\     
\bottomrule
\end{tabular}
\label{table:PL}
\end{minipage}
\begin{tablenotes}
      \item \tiny{All data are expressed as median values. Legend: Thr, R threshold; $\lambda$, $\alpha$ model parameters; $x_{min}$, lower bound for model distribution; TL, length of the tail; KS, Kolgomorov-Smirnov statistic; \emph{p}-value, plausibility of the model; TL$_{r}$, proportion of non-zero nodes in the tail.}
\end{tablenotes}
\end{table}

\section{Results}
Power-law testing was performed on Matlab (Mathworks Inc., USA) using \cite{Clauset2009}. Further, for testing exponential, log-normal and generalized Pareto models we adapted the framework provided in \cite{Clauset2009} to include these competing hypothesis. For each subject, we analyzed thresholds in the range $-0.7$ to $0.8$, with $0.1$ increments. The parametric \emph{goodness-of-fit} test was conducted over 1,000 repetitions, ensuring precision of \emph{p}-value up to two decimal digits.

\begin{figure}[ht!]
\centering
\includegraphics[scale=.75]{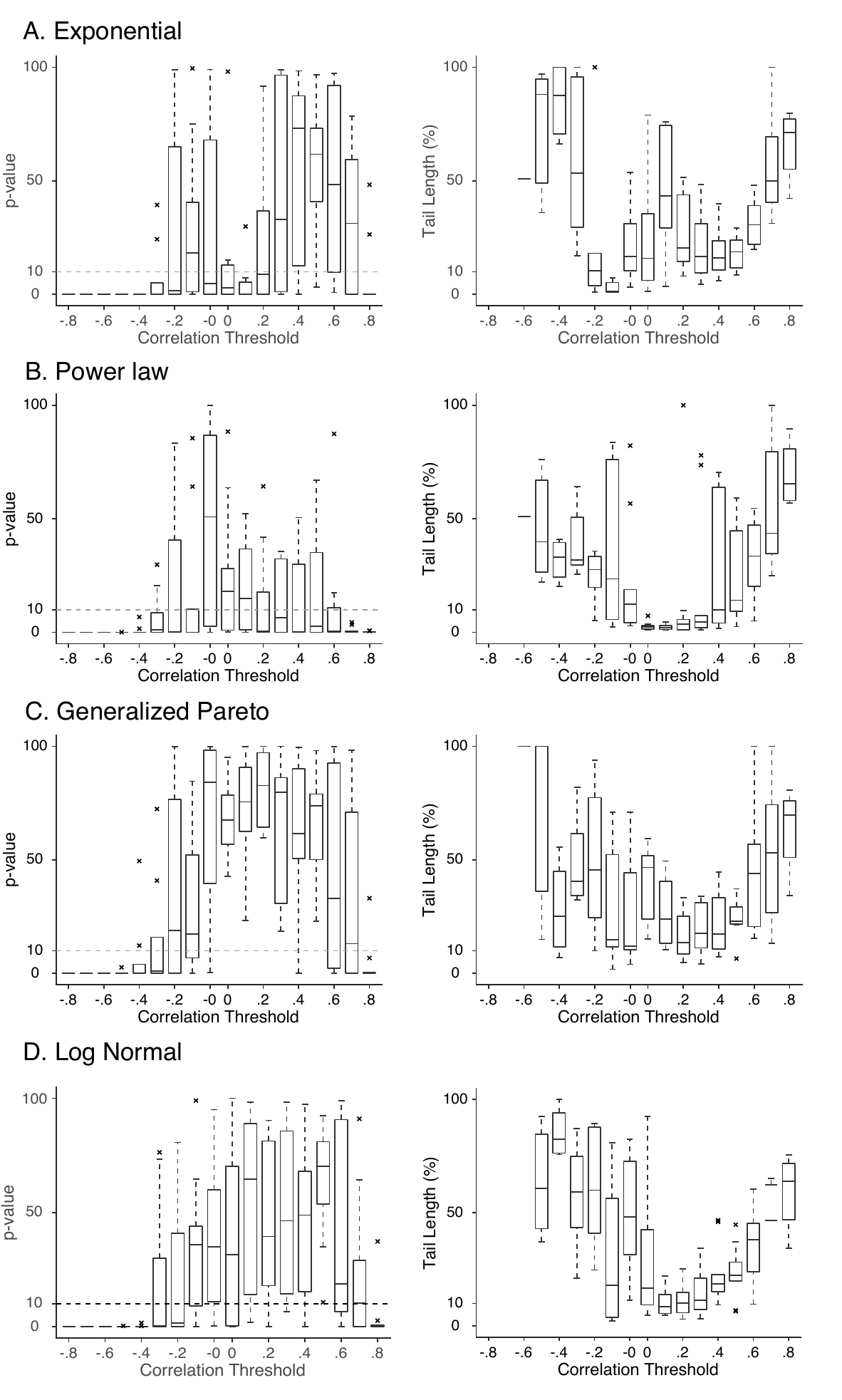}
\caption{Population averaged goodness-of-fit tests (left) and percentage of the tail of the distribution  explained by the model (right) across different thresholds for each of the four distributions. Horizontal dashed lines in the boxplots indicate the acceptance criteria for a model to be considered plausible ($p$-value$>10\%$). The central mark is the median, the edges of the boxes are the 25$^{th}$ and 75$^{th}$ percentiles. Asterisks correspond to outliers. \label{fig:02}}
\end{figure}

Our results are summarized in tables~\ref{table:EXP} and \ref{table:GP}, respectively. An hypothesis is considered plausible if the \emph{p}-value is larger than 0.1. 
Averaging over subjects, the \emph{p}-values indicate that the power law hypothesis is rejected in the 83.3$\%$ of the analyzed thresholds. Instead,  61.1$\%$ of the examined thresholds are consistent with a generalized Pareto hypothesis, 55.6$\%$ with a log-normal hypothesis and in 33.3$\%$ of the cases with the exponential hypothesis, with several of these thresholds passing multiple tests. Median \emph{p}-values are consistently larger for the generalized Pareto hypothesis (figure~\ref{fig:02}). 

For each threshold examined, we then perform log-likelihood ratio tests to  check which one among the consistent models is the most plausible in describing the empirical data.
For all the positive thresholds up to 0.7 the evidence strongly goes in favor of the generalized Pareto distribution. Overall, the generalized Pareto model is outperforming the other candidate models in 41$\%$ of the examined cases (all subjects and all thresholds). In a 13$\%$ of the comparisons the log-normal distribution resulted in a better fit, 3$\%$ were better fitted by an exponential model, 2$\%$ by a power law, whereas the remaining 41$\%$ could not be explained by any model (due to insufficient data points and extreme thresholds).

\begin{figure}[tb!]
\centering
\includegraphics[scale=0.85]{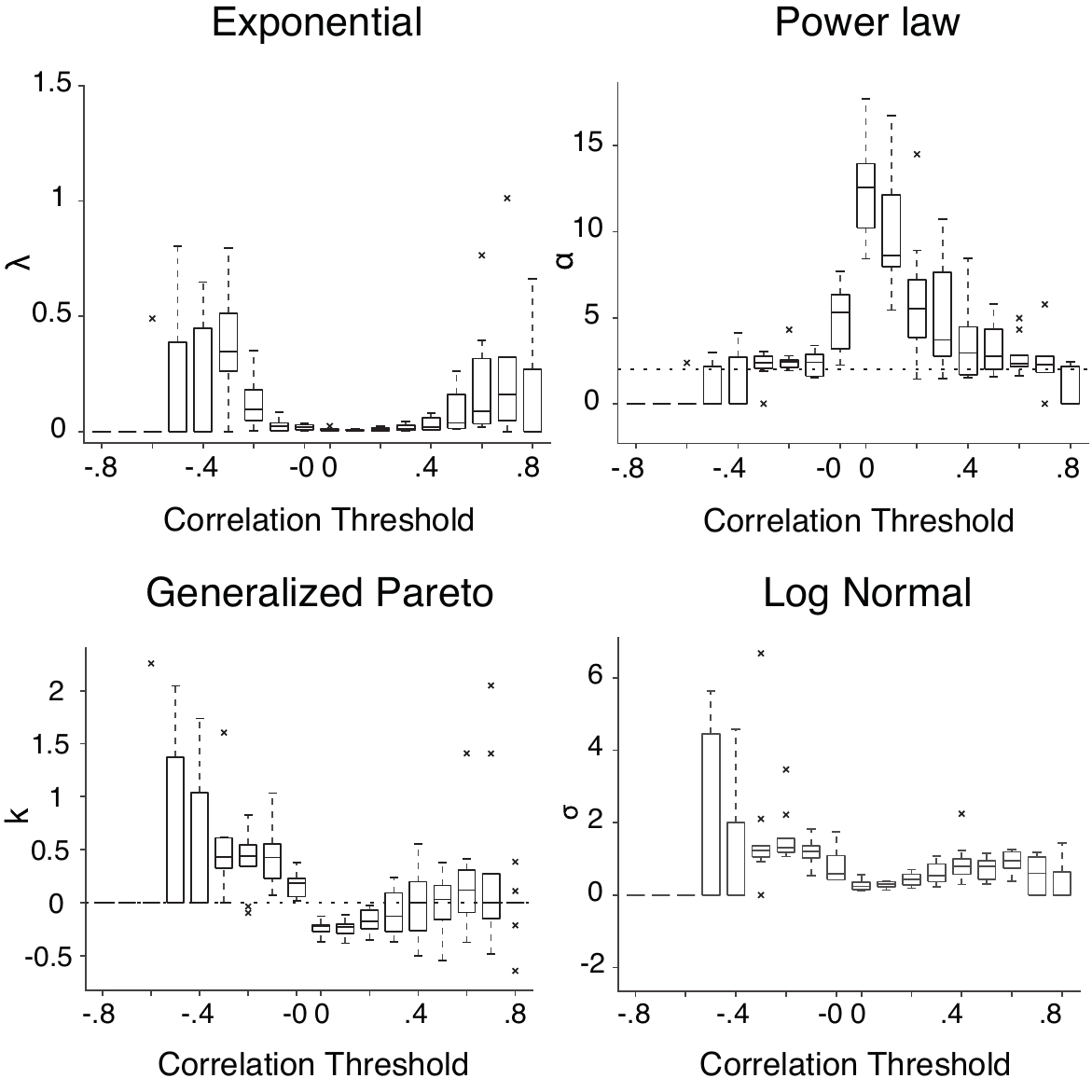}
\caption{Population averaged estimates of $\lambda$ (top-left), $\alpha$ (top-right), \emph{k} (bottom-left) and $\sigma$ (bottor-right) model parameters for the four tested distributions across different thresholds.\label{fig:03}}
\end{figure}

For several positive thresholds the $k$ parameter is equal or close to zero, thus  approaching an exponential distribution, whereas for other thresholds in the positive range, the generalized Pareto model passes with negative $k$, meaning a suppressed tail (figure~\ref{fig:03}). 

\begin{table}[t!]                           
\centering 
\caption{Fit results of generalized Pareto distribution ($p(x) = \frac{1}{\sigma} (1+k\frac{x-x_{min}}{\sigma})^{-1-\frac{1}{k}}$) and of log-normal distribution ($C\frac{1}{x}exp[\frac{-(ln(x)-\mu)^{2}}{2\sigma^{2}}]$).}
\begin{minipage}[t]{.35\linewidth}
\centering
\tiny 
\caption*{\hspace*{15mm}Generalized Pareto}
\begin{tabular}{lrrrrrccc}
\toprule
\multicolumn{1}{c}{Thr} & \multicolumn{1}{c}{$k$} & \multicolumn{1}{c}{$\sigma$} & \multicolumn{1}{c}{$x_{min}$} & \multicolumn{1}{c}{TL} & \multicolumn{1}{c}{KS} &  \multicolumn{1}{c}{p}-value & \multicolumn{1}{c}{TL$_r$} \\
\midrule
$+0.8$ & 0.000 & 0.000 & 0.000 & 0.0 & 0.000 & 0.0000 & 0.696 \\  
$+0.7$ & 0.000 & 3.663 & 1.539 & 58.5 & 0.049 & 0.1315 & 0.531  \\   
$+0.6$ & 0.118 & 11.094 & 2.943 & 172.5 & 0.048 & 0.3305 & 0.440  \\  
$+0.5$ & 0.031 & 28.656 & 20.072 & 338.5 & 0.034 & 0.7380 & 0.229 \\  
$+0.4$ & 0.001 & 62.035 & 34.191 & 372.5 & 0.027 & 0.6155 & 0.174 \\  
$+0.3$ & -0.128 & 102.619 & 85.215 & 638.5 & 0.021 & 0.7975 & 0.176\\
$+0.2$ & -0.178 & 152.365 & 136.945 & 938.5 & 0.014 & 0.8275 & 0.135  \\  
$+0.1$ & -0.226 & 241.640 & 201.475 & 2387.0 & 0.010 & 0.7555 & 0.238\\
$+0.0$ & -0.224 & 218.500 & 380.050 & 4681.5 & 0.009 & 0.6750 & 0.467 \\ 
$-0.0$ & 0.182 & 32.619 & 127.480 & 1195.5 & 0.012 & 0.8425 & 0.119 \\
$-0.1$ & 0.426 & 12.883 & 9.690 & 1469.0 & 0.020 & 0.1725 & 0.149 \\   
$-0.2$ & 0.439 & 2.549 & 0.763 & 820.5 & 0.030 & 0.1890 & 0.457 \\     
$-0.3$ & 0.433 & 0.996 & 0.649 & 101.5 & 0.062 & 0.0105 & 0.405 \\
$-0.4$ & 0.000 & 0.000 & 0.000 & 0.0 & 0.000 & 0.0000 & 0.253 \\              
$-0.5$ & 0.000 & 0.000 & 0.000 & 0.0 & 0.000 & 0.0000 & 1.000 \\     
$-0.6$ & 0.000 & 0.000 & 0.000 & 0.0 & 0.000 & 0.0000 & 1.000 \\  
$-0.7$ & 0.000 & 0.000 & 0.000 & 0.0 & 0.000 & 0.0000 & - \\    
$-0.8$ & 0.000 & 0.000 & 0.000 & 0.0 & 0.000 & 0.0000 & - \\ 
\bottomrule
\end{tabular}                               
\label{table:GP} 
\end{minipage}
\begin{minipage}[t]{.45\linewidth}
\centering
\tiny
\caption*{Log normal}
\begin{tabular}{lrrrrccc}
\toprule
\multicolumn{1}{c}{Thr} & \multicolumn{1}{c}{$\mu$} & \multicolumn{1}{c}{$\sigma$} & \multicolumn{1}{c}{$x_{min}$} & \multicolumn{1}{c}{TL} & \multicolumn{1}{c}{KS} &  \multicolumn{1}{c}{p}-value & \multicolumn{1}{c}{TL$_r$}\\
\midrule
$+0.8$ & 0.000 & 0.000 & 0.000 & 0.0 & 0.000 & 0.0000  & 0.639 \\   
$+0.7$ & 0.618 & 0.599 & 1.141 & 58.0 & 0.039 & 0.1033  & 0.466 \\
$+0.6$ & 1.804 & 0.957 & 2.589 & 110.5 & 0.040 & 0.1867  & 0.381 \\        
$+0.5$ & 3.363 & 0.794 & 15.167 & 246.5 & 0.033 & 0.7017  & 0.223 \\ 
$+0.4$ & 3.459 & 0.799 & 24.044 & 543.5 & 0.027 & 0.4883  & 0.186 \\
$+0.3$ & 4.809 & 0.535 & 72.627 & 525.0 & 0.023 & 0.4633  & 0.114 \\                  
$+0.2$ & 5.754 & 0.434 & 209.015 & 759.5 & 0.023 & 0.3950  & 0.100 \\           
$+0.1$ & 6.143 & 0.299 & 418.120 & 852.0 & 0.019 & 0.6450  & 0.085 \\                 
$+0.0$ & 6.299 & 0.236 & 472.295 & 1684.5 & 0.015 & 0.3150  & 0.168 \\                
$-0.0$ & 4.298 & 0.588 & 81.662 & 4837.0 & 0.010 & 0.3500  & 0.482 \\           
$-0.1$ & 2.176 & 1.197 & 7.105 & 1764.0 & 0.019 & 0.3600  & 0.180 \\   
$-0.2$ & 0.486 & 1.317 & 0.610 & 879.0 & 0.034 & 0.0167  & 0.599 \\          
$-0.3$ & -0.243 & 1.237 & 0.543 & 106.0 & 0.072 & 0.0050  & 0.592 \\        
$-0.4$ & 0.000 & 0.000 & 0.000 & 0.0 & 0.000 & 0.0000  & 0.825 \\           
$-0.5$ & 0.000 & 0.000 & 0.000 & 0.0 & 0.000 & 0.0000  & 0.607 \\                  
$-0.6$ & 0.000 & 0.000 & 0.000 & 0.0 & 0.000 & 0.0000  & -Inf \\     
$-0.7$ & 0.000 & 0.000 & 0.000 & 0.0 & 0.000 & 0.0000  & - \\  
$-0.8$ & 0.000 & 0.000 & 0.000 & 0.0 & 0.000 & 0.0000  & - \\           
\bottomrule
\end{tabular}
\label{table:LN}
\end{minipage}
\begin{tablenotes}
      \item \tiny{All data are expressed as median values. Legend: Thr, R threshold; $k$, $\sigma$, $\mu$ model parameters; $x_{min}$, lower bound for model distribution; TL, length of the tail; KS, Kolgomorov-Smirnov statistic; \emph{p}-value, plausibility of the model; TL$_{r}$, proportion of non-zero nodes in the tail.}
\end{tablenotes}
\end{table}

\section{Discussion}
In this study we sought to systematically analyze scaling properties of human brain functional networks in the resting state, obtained from high-resolution fMRI data. We constructed networks of 10,000 nodes. Our analysis took into account actual weighted degree distributions from the data and we scanned through the full range of positive as well as negative correlation thresholds. For model selection, we imposed a criterion of \emph{p}-value $> 0.1$ and we conducted a log-likelihood ratio test among the different hypothesis. 

We have shown that the degree distribution of the nodes does not follow a scale-free topology, as reported in~\cite{Eguiluz2005}. The power law hypothesis is strongly rejected in the majority of the thresholds we examined. Indeed, it is the generalized Pareto distribution that is consistently preferable to the competing models for most of the thresholds. 

These results suggest that after taking into account continuously weighted rather than binary networks, the dynamics of brain functional networks might not be governed by as many ultra-high degree hubs as a typical scale-free network might suggest. This bodes well for real brain networks when considering resilience to attacks, compared to their scale-free counterparts. For future work, we intend to test  whether these distributions hold for different network resolutions and parcellations. Moreover, it would be interesting to see how these results compare to the ``core-periphery'' organization of brain structural networks  \cite{harriger2012rich,van2011rich}, which shows a preference for a distributed core, rather than few ultra-high degree hubs.  

\subsubsection*{\small{Acknowledgments.}} \small{The work has been supported by the European Research Council under the EU’s \emph{7th Framework Programme} (FP7/2007-2013)/ERC grant agreement no.~341196 to P.~Verschure.
Data were provided [in part] by the Human Connectome Project, WU-Minn Consortium (Principal Investigators: David Van Essen and Kamil Ugurbil; 1U54MH091657) funded by the 16~NIH Institutes and Centers that support the NIH~Blueprint for Neuroscience Research; and by the McDonnell Center for Systems Neuroscience at Washington University.}

\bibliographystyle{splncs03}
\bibliography{ICANN}

\begin{thebibliography}{10}
\providecommand{\url}[1]{\texttt{#1}}
\providecommand{\urlprefix}{URL }

\bibitem{Achard2006}
Achard, S., Salvador, R., Whitcher, B., Suckling, J., Bullmore, E.: {A
  resilient, low-frequency, small-world human brain functional network with
  highly connected association cortical hubs.} The Journal of Neuroscience :
  the official journal of the Society for Neuroscience  26(1),  63--72 (2006)

\bibitem{Albert2000}
Albert, R., Jeong, H., Barab{\'a}si, A.L.: Error and attack tolerance of
  complex networks. nature  406(6794),  378--382 (2000)

\bibitem{ECMS}
Arsiwalla, X.D., Betella, A., Mart{\'\i}nez, E., Omedas, P., Zucca, R.,
  Verschure, P.: The dynamic connectome: a tool for large scale 3d
  reconstruction of brain activity in real time. In: 27th European Conference
  on Modeling and Simulation. ECMS W. Rekdalsbakken, R. Bye, H. Zhang eds.,
  ECMS W. Rekdalsbakken, R. Bye, H. Zhang eds., Alesund (Norway) (2013)

\bibitem{2015connectomics}
Arsiwalla, X.D., Dalmazzo, D., Zucca, R., Betella, A., Brandi, S., Martinez,
  E., Omedas, P., Verschure, P.: Connectomics to semantomics: Addressing the
  brain's big data challenge. Procedia Computer Science  53,  48--55 (2015)

\bibitem{2015network}
Arsiwalla, X.D., Zucca, R., Betella, A., Martinez, E., Dalmazzo, D., Omedas,
  P., Deco, G., Verschure, P.: Network dynamics with brainx3: A large-scale
  simulation of the human brain network with real-time interaction. Frontiers
  in Neuroinformatics  9(2) (2015)

\bibitem{Clauset2009}
Clauset, A., Shalizi, C., Newman, M.: {Power-law distributions in empirical
  data}. SIAM review  (2009)

\bibitem{Eguiluz2005}
Egu{\'{\i}}luz, V.M., Chialvo, D.R., Cecchi, G.a., Baliki, M., Apkarian, a.V.:
  {Scale-Free Brain Functional Networks}. Physical Review Letters  94(1),
  018102 (2005)

\bibitem{Fornito2010}
Fornito, A., Zalesky, A., Bullmore, E.T.: {Network scaling effects in graph
  analytic studies of human resting-state fMRI data}. Frontiers in Systems
  Neuroscience  4, ~22 (2010)

\bibitem{harriger2012rich}
Harriger, L., Van Den~Heuvel, M.P., Sporns, O.: Rich club organization of
  macaque cerebral cortex and its role in network communication. PloS one
  7(9),  e46497 (2012)

\bibitem{Hayasaka2010}
Hayasaka, S., Laurienti, P.J.: {Comparison of characteristics between
  region-and voxel-based network analyses in resting-state fMRI data}.
  NeuroImage  50(2),  499--508 (2010)

\bibitem{VandenHeuvel2008}
van~den Heuvel, M.P., Stam, C.J., Boersma, M., {Hulshoff Pol}, H.E.:
  {Small-world and scale-free organization of voxel-based resting-state
  functional connectivity in the human brain}. NeuroImage  43(3),  528--539
  (2008)

\bibitem{van2011rich}
van~den Heuvel, M.P., Sporns, O.: Rich-club organization of the human
  connectome. The Journal of Neuroscience  31(44),  15775--15786 (2011)

\bibitem{vanEssen2013}
{Van Essen}, D.C., Smith, S.M., Barch, D.M., Behrens, T.E.J., Yacoub, E.,
  Ugurbil, K.: {The WU-Minn Human Connectome Project: An overview}. NeuroImage
  80,  62--79 (2013)

\end{thebibliography}
\end{document}